\documentstyle[12pt]{article}
\input amssym.def
\input amssym.tex

\newcommand{\wx}{\widetilde{x}}
\newcommand{\wip}{\widetilde{p}}
\newcommand{\wa}{\widetilde{a}}
\newcommand{\wb}{\widetilde{b}}
\newcommand{\wc}{\widetilde{c}}
\newcommand{\wid}{\widetilde{d}}

\begin{document}

\newcommand{\eto}[1]{\mbox{$e^{\displaystyle #1}$}} 
\newcommand{\nm}{\!-\!}
\newcommand{\np}{\!+\!}
\newcommand{\ueto}[1]{\raisebox{3mm}{$#1$}}

\def\m@th{\mathsurround=0pt}
\mathchardef\bracell="0365 
\def\upbrall{$\m@th\bracell$}
\def\undertilde#1{\mathop{\vtop{\ialign{##\crcr
    $\hfil\displaystyle{#1}\hfil$\crcr
     \noalign
     {\kern1.5pt\nointerlineskip}
     \upbrall\crcr\noalign{\kern1pt
   }}}}\limits}

\begin{titlepage}
 {\LARGE
  \begin{center} 
A collection of integrable systems \\of the Toda type in continuous and
discrete time, \\with $2\times 2$ Lax representations   \end{center}
 }

\vspace{1.5cm}

\begin{flushleft}{\large Yuri B. SURIS}\end{flushleft} \vspace{1.0cm}
Centre for Complex Systems and Visualization, University of Bremen,\\
Universit\"atsallee 29, 28359 Bremen, Germany\\
e-mail: suris @ cevis.uni-bremen.de 

\vspace{2.0cm}

{\small {\bf Abstract.} A fairly complete list of Toda--like integrable
lattice systems, both in the continuous and discrete time, is given.
For each system the Newtonian, Lagrangian and Hamiltonian formulations
are presented, as well as the $2\times 2$ Lax representation and $r$--matrix
structure. The material is given in the ''no comment'' style, in particular, 
all proofs are omitted.}
\end{titlepage}

\setcounter{section}{-1}
\section{Introduction}
This paper contains almost exclusively a large number of formulas presented
in the ''no comment''--style. I hope that they will speak for themselves
and tell once more the story on how much contain two simple systems:
the Toda lattice and the relativistic Toda lattice. I have only tried to
collect many closely related results in the form that makes the interrelations
quite transparent. All proofs are omitted, because they consist of a direct
verification. 

I present a large number of integrable Newtonian systems, both in the 
continuous and discrete time, and give for all of them the equivalent 
Lagrangian and, for the continuous time systems, also Hamiltonian formulations 
(generalities are given in the Sect. 1), and also present $2\times 2$ 
Lax representations. All these systems are reparametrizations of the 
Toda and the relativistic Toda lattices or their discrete time analogs
(reminded in the Sect. 2, 3). For these systems the ''big'' Lax representations 
are well known (they include $N\times N$ matrices for the $N$--particle
lattices). So, the corresponding reparametrizations immediately imply the
''big'' Lax representations for all our systems. In principle, the
transition from the ''big'' Lax representation to the $2\times 2$ one is
well understood, at least at the formal level: they correspond to two
different ways to represent a spectral problem connected with the second 
order linear difference operator. However, such transition is not
uniquely defined, it may be performed in infinitely many different ways
leading to different but gauge equivalent $2\times 2$ Lax matrices. The
problem is to find the gauge leading to the most nice Lax matrices.
In particular, it is usualy required that the matrix $L_k$ depends only
on $x_k$, $p_k$, the coordinates and momenta of the $k$th particle.
To find such gauge is sometimes a nontrivial task. I give here a collection
of Lax representations having this property.

The last Section contains several historical remarks.

\setcounter{equation}{0}
\section{Newtonian, Lagrangian, and Hamiltonian \newline systems 
in continuous and discrete time}
In what follows we consider continuous time lattice systems in the
Newtonian form:
\begin{equation}
\ddot{x}_k=F_k(x,\dot{x})
\end{equation}
They all may be put into the Euler--Lagrange form
\begin{equation}\label{New}
\frac{d}{dt}\frac{\partial{\cal L}(x,\dot{x})}{\partial \dot{x}_k}+
\frac{\partial{\cal L}(x,\dot{x})}{\partial x_k}=0
\end{equation}
We will call the Lagrangian formulation of such equations the system
\begin{equation}\label{Lagr}
\left\{\begin{array}{l}
p_k=\partial{\cal L}(x,\dot{x})/\partial\dot{x}_k \\ \\
\dot{p}_k=-\partial{\cal L}(x,\dot{x})/\partial x_k
\end{array}\right.
\end{equation}
The Hamiltonian formulation of these equations:
\begin{equation}\label{Ham}
\left\{\begin{array}{l}
\dot{x}_k=\partial H(x,p)/\partial p_k \\ \\
\dot{p}_k=-\partial H(x,p)/\partial x_k
\end{array}\right.
\end{equation}
The well--known Legendre transform gives a relation between the Lagrange and 
the Hamilton functions:
\begin{equation}\label{L to H}
H(x,p)=\sum_{k} \dot{x}_kp_k-{\cal L}(x,\dot{x})
\end{equation}
A discrete time analog of the Euler--Lagrange equations (\ref{New}):
\begin{equation}\label{dNew}
\frac{\partial}{\partial x_k}\Big(\Lambda(\wx,x)+\Lambda(x,\undertilde{x})\Big)
=0
\end{equation}
We will call the Lagrangian formulation of such equations the system
\begin{equation}\label{dLagr}
\left\{\begin{array}{l}
p_k=-\partial\Lambda(\wx,x)/\partial x_k \\ \\
\wip_k=\partial\Lambda(\wx,x)/\partial \wx_k
\end{array}\right.
\end{equation}
The Hamiltonian formulation in the discrete time case is not well--defined.
However, for the systems related to the relativistic Toda hierarchy, we will
find nice approximations of the Hamiltonian equations of motion (\ref{Ham}).

\setcounter{equation}{0}
\section{Simplest flow of the Toda hierarchy\newline
and its bi--Hamiltonian structure}
The simplest flow of the Toda hierarchy (hereafter denoted by TL) is:
\begin{equation}\label{TL}
\dot{a}_k=a_k(b_{k+1}-b_k), \qquad \dot{b}_k=a_k-a_{k-1}
\end{equation}
Its discretization (called hereafter dTL): 
\begin{equation}\label{dTL}
\wa_k=a_k\,\frac{\beta_{k+1}}{\beta_k},\quad
\wb_k=b_k+h\left(\frac{a_k}{\beta_k}-\frac{a_{k-1}}{\beta_{k-1}}\right) 
\end{equation}
where $\beta_k=\beta_k(a,b)$ are defined as a unique set of functions 
satisfying the relations
\begin{equation}\label{dTL recur}
\beta_k=1+hb_k-\frac{h^2a_{k-1}}{\beta_{k-1}}=1+hb_k+O(h^2)
\end{equation}
Both the TL and the dTL are bi--Hamiltonian. The first (''linear'') invariant 
Poisson bracket:
\begin{equation}\label{TL l br}
\{a_k,b_k\}_1=-\{a_k,b_{k+1}\}_1=a_k
\end{equation}
The second (''quadratic'') one:
\begin{equation}\label{TL q br}
\begin{array}{cclcccl}
\{a_k,a_{k+1}\}_2 & = & -a_{k+1}a_k,& \quad & \{b_k,b_{k+1}\}_2 & = & -a_k\\
\{a_k,b_k\}_2 & = & a_kb_k,&  \quad & \{a_k,b_{k+1}\}_2 & = & -a_kb_{k+1}
\end{array}
\end{equation}
The Hamilton functions generating the flow TL in these brackets:
\begin{equation}\label{TL H}
H_1(a,b)=\frac{1}{2}\sum_{k}b_k^2+\sum_{k}a_k,\qquad
H_2(a,b)=\sum_{k} b_k
\end{equation}

\setcounter{equation}{0}
\section{Simplest flows of the relativistic Toda hierarchy\newline
and their bi--Hamiltonian structure}

The first flow of the relativistic Toda hierarchy (denoted hereafter RTL+):
\begin{equation}\label{RTL+}
\dot{d}_k=d_k(c_k-c_{k-1}), \qquad
\dot{c}_k=c_k(d_{k+1}+c_{k+1}-d_k-c_{k-1})
\end{equation}
The second flow of the relativistic Toda hierarchy (denoted hereafter RTL$-$):
\begin{equation}\label{RTL-}
\dot{d}_k=d_k\left(\frac{c_k}{d_kd_{k+1}}-
\frac{c_{k-1}}{d_{k-1}d_k}\right), \quad
\dot{c}_k=c_k\left(\frac{1}{d_k}-\frac{1}{d_{k+1}}\right)
\end{equation}

An integrable discretization of the flow RTL+ (denoted hereafter dRTL+):
\begin{equation}\label{dRTL+}
\widetilde{d}_k=d_k\,\frac{{\goth a}_{k+1}-hd_{k+1}}{{\goth a}_k-hd_k},\qquad
\widetilde{c}_k=c_k\,\frac{{\goth a}_{k+1}+hc_{k+1}}{{\goth a}_k+hc_k}
\end{equation}
where ${\goth a}_k={\goth a}_k(c,d)$ is defined as a unique set of functions 
satisfying the relations
\begin{equation}\label{dRTL+ recur}
{\goth a}_k=1+hd_k+\frac{hc_{k-1}}{{\goth a}_{k-1}}=1+h(d_k+c_{k-1})+O(h^2)
\end{equation}

An integrable discretization of the flow RTL-- (denoted hereafter dRTL$-$):
\begin{equation}\label{dRTL-}
\widetilde{d}_k=
d_{k+1}\,\frac{d_k-h{\goth d}_{k-1}}{d_{k+1}-h{\goth d}_k}, \quad
\widetilde{c}_k=
c_{k+1}\,\frac{c_k+h{\goth d}_k}{c_{k+1}+h{\goth d}_{k+1}}
\end{equation}
where ${\goth d}_k={\goth d}_k(c,d)$ is defined as a unique set of functions 
satisfying the relations
\begin{equation}\label{dRTL- recur}
{\goth d}_k=\frac{c_k}{d_k-h-h{\goth d}_{k-1}}=\frac{c_k}{d_k}+O(h)
\end{equation}

An integrable difference equation (hereafter called explicit dRTL):
\begin{equation}\label{edRTL}
\wid_k=d_{k-1}\,\frac{d_k+c_k}{d_{k-1}+c_{k-1}},\quad
\wc_k=c_k\,\frac{d_{k+1}+c_{k+1}}{d_k+c_k}
\end{equation}
This map is not close to the identity and therefore cannot serve as a 
discretization of some flow of the RTL hierarchy. However, it becomes related 
to the flow TL, if 
\begin{equation}
d_k\approx 1+hb_k,\qquad c_k\approx h^2a_k
\end{equation}
(see Sect. 7 for details).

The flows RTL$\pm$ and the maps dRTL$\pm$ and explicit dRTL are
bi--Hamiltonian. The first (linear) invariant Poisson bracket:
\begin{equation}\label{RTL l br}
\{c_k,d_{k+1}\}_1=-c_k, \quad \{c_k,d_k\}_1=c_k,\quad \{d_k,d_{k+1}\}_1=c_k
\end{equation}
(only the non--vanishing brackets are written down), and Hamilton functions
generating the flows (\ref{RTL+}), (\ref{RTL-}) in this bracket:
\begin{eqnarray}
H^{(+)}_1(c,d) & = & \frac{1}{2}\sum_{k}(d_k+c_{k-1})^2+
\sum_{k}(d_k+c_{k-1})c_k \label{RTL+ H1}\\
H^{(-)}_1(c,d) & = & -\sum_{k}\log(d_k) \label{RTL- H1}
\end{eqnarray}
The second (quadratic) invariant Poisson bracket:
\begin{equation}\label{RTL q br}
\{c_k,c_{k+1}\}_2=-c_kc_{k+1}, \quad \{c_k,d_{k+1}\}_2=-c_kd_{k+1}, \quad
\{c_k,d_k\}_2=c_kd_k
\end{equation}
the corresponding Hamilton functions:
\begin{eqnarray}
H^{(+)}_2(c,d) & = & \sum_{k} (d_k+c_k)\label{RTL+ H2}\\
H^{(-)}_2(c,d) & = & \sum_{k}\frac{d_k+c_k}{d_kd_{k+1}}\label{RTL- H2}
\end{eqnarray}

\setcounter{equation}{0}
\section{Systems TL and dTL: 
\newline Parametrization of the linear bracket}

Canonical parametrization of the bracket $\{\cdot,\cdot\}_1$:
\begin{equation}\label{TL l par}
a_k=\eto{x_{k+1}\nm x_k},\qquad b_k=p_k
\end{equation}
The resulting system is the {\it conventional Toda lattice}.

\subsection{The flow TL}

Hamilton function:
\begin{equation}\label{TL H1 in xp}
H(x,p)=H_1(a,b)=\frac{1}{2}\sum_{k}p_k^2+\sum_{k}\eto{x_k\nm x_{k-1}}
\end{equation}
Hamiltonian equations of motion:
\begin{equation}
\left\{\begin{array}{l}
\dot{x}_k = p_k\\ \\
\dot{p}_k = \eto{x_{k+1}\nm x_k}-\eto{x_k\nm x_{k-1}}
\end{array}\right.
\end{equation}
Newtonian equations of motion:
\begin{equation}\label{Toda New}
\ddot{x}_k=\eto{x_{k+1}\nm x_k}-\eto{x_k\nm x_{k-1}}
\end{equation}

\subsection{The map dTL}

Lagrangian form of equations of motion:
\begin{equation}
\left\{\begin{array}{l}
hp_k=\eto{\wx_k\nm x_k}-1+h^2\eto{x_k\nm\wx_{k-1}}\\ \\
h\wip_k = \eto{\wx_k\nm x_k}-1+h^2\eto{x_{k+1}\nm\wx_k}
\end{array}\right.
\end{equation}
Newtonian form of equations of motion:
\begin{equation}
e\ueto{\wx_k\nm x_k}-e\ueto{x_k\nm\undertilde{x_k}}=
h^2\bigg(e\ueto{\,\undertilde{x_{k+1}}\nm x_k}-
e\ueto{x_k\nm\wx_{k-1}}\bigg)
\end{equation}

\subsection{Lax representations and $r$--matrix structure}
The both systems above allow a Lax representation with the Lax matrix
\begin{equation}\label{Toda L}
L_k=\left(\begin{array}{cc}
p_k+\lambda & \eto{x_k}\\ \\-\eto{-x_k} & 0\end{array}\right)
\end{equation}
This matrix satisfies the $r$--matrix ansatz
\begin{equation}\label{rmat ans 1}
\{L_k(\lambda)\stackrel{\otimes}{,}L_j(\mu)\}=\left[\frac{P}{\lambda-\mu},
L_k(\lambda)\otimes L_k(\mu)\right]\delta_{kj}
\end{equation}
with
\begin{equation}
P=\left(\begin{array}{cccc} 1 & 0 & 0 & 0 \\ 0 & 0 & 1 & 0 \\
0 & 1 & 0 & 0 \\ 0 & 0 & 0 & 1 \end{array}\right)
\end{equation}
which assures the complete integrabilty.

The Lax representation for the flow TL:
\begin{equation}\label{Lax}
\dot{L}_k=M_{k+1}L_k-L_kM_k
\end{equation}
with
\begin{equation}\label{Toda M}
M_k=\left(\begin{array}{cc}
-\lambda & -\eto{x_k}\\ \\ \eto{-x_{k-1}} & 0\end{array}\right)
\end{equation}
The Lax representation for the map dTL:
\begin{equation}\label{dLax}
\widetilde{L}_kV_k=V_{k+1}L_k
\end{equation}
with 
\begin{equation}
V_k=\left(\begin{array}{cc}
1-h\lambda-h^2\eto{x_k\nm\wx_{k-1}} & -h\eto{x_k}\\ \\
h\eto{-\wx_{k-1}} & 1\end{array}\right)
\end{equation}

\setcounter{equation}{0}
\section{Systems TL and dTL: 
\newline Parametrization of the quadratic bracket} 
 
Canonical parametrization of the bracket $\{\cdot,\cdot\}_2$:
\begin{equation}\label{TL q par}
a_k=\eto{x_{k+1}\nm x_k\np p_k},\qquad b_k=\eto{p_k}+\eto{x_k\nm x_{k-1}}.
\end{equation}
The resulting system is the {\it modified Toda lattice}.

\subsection{The flow TL}

Hamilton function:
\begin{equation}\label{TL H2 in xp}
H(x,p)=H_2(a,b)=\sum_{k}\eto{p_k}+\sum_{k}\eto{x_k\nm x_{k-1}}
\end{equation}
Hamiltonian equations of motion:
\begin{equation}\left\{
\begin{array}{l}
\dot{x}_k = \eto{p_k}\\ \\
\dot{p}_k = \eto{x_{k+1}\nm x_k}-\eto{x_k\nm x_{k-1}}
\end{array}\right.
\end{equation}
Newtonian equations of motion:
\begin{equation}\label{mToda New}
\ddot{x}_k=\dot{x}_k\left(\eto{x_{k+1}\nm x_k}-\eto{x_k\nm x_{k-1}}\right)
\end{equation}

\subsection{The map dTL}

Lagrangian form of equations of motion.
\begin{equation}
\left\{\begin{array}{l}
h\eto{p_k} =\left(\eto{\wx_k\nm x_k}-1\right)\,
\left(1+h\eto{x_k\nm\wx_{k-1}}\right)\\ \\
h\eto{\wip_k} = \left(\eto{\wx_k\nm x_k}-1\right)\,
\left(1+h\eto{x_{k+1}\nm\wx_k}\right)
\end{array}\right.
\end{equation}
Newtonian form of equations of motion:
\begin{equation}\label{dmToda New}
\frac{\Big(\eto{\wx_k\nm x_k}-1\Big)}
{\Big(e\ueto{x_k\nm\undertilde{x_k}}-1\Big)}=
\frac{\bigg(1+he\ueto{\,\undertilde{x_{k+1}}\nm x_k}\bigg)}
{\bigg(1+h\eto{x_k\nm\wx_{k-1}}\bigg)}
\end{equation}

\subsection{Lax representations and $r$--matrix structure}
Lax representations for the both systems above include the Lax matrix
\begin{equation}\label{mToda L}
L_k=\left(\begin{array}{cc}
\lambda\eto{p_k}-\lambda^{-1} & \eto{x_k}\\ \\
-\eto{-x_k} & \lambda\end{array}\right)
\end{equation}
satisfying the $r$--matrix ansatz
\begin{equation}\label{rmat 2 ans}
\{L_k(\lambda)\stackrel{\otimes}{,}L_j(\mu)\}=\Big[r(\lambda,\mu)\,,\,
L_k(\lambda)\otimes L_k(\mu)\Big]\delta_{kj}
\end{equation}
with
\begin{equation}\label{r}
r(\lambda,\mu)=\left(\begin{array}{cccc} 
\frac{1}{2}\frac{\lambda^2+\mu^2}{\lambda^2-\mu^2} & 0 & 0 & 0 \\ 
0 & -\frac{1}{2} & \frac{\lambda\mu}{\lambda^2-\mu^2} & 0 \\
0 & \frac{\lambda\mu}{\lambda^2-\mu^2} & \frac{1}{2} & 0 \\ 
0 & 0 & 0 & \frac{1}{2}\frac{\lambda^2+\mu^2}{\lambda^2-\mu^2} 
\end{array}\right)
\end{equation}
The Lax representation for the flow TL: (\ref{Lax}) with
\begin{equation}\label{mToda M}
M_k=\left(\begin{array}{cc}
\lambda^{-2}+\eto{x_k\nm x_{k-1}} & -\lambda^{-1}\eto{x_k}\\ \\ 
\lambda^{-1}\eto{-x_{k-1}} & 0\end{array}\right)
\end{equation}
The Lax representation for the map dTL: (\ref{dLax}) with
\begin{equation}
V_k=\left(\begin{array}{cc}
1+h\lambda^{-2}+h\eto{x_k\nm\wx_{k-1}} & -h\lambda^{-1}\eto{x_k}\\ \\
h\lambda^{-1}\eto{-\wx_{k-1}} & 1\end{array}\right)
\end{equation}

\setcounter{equation}{0}
\section{Systems TL and dTL:
\newline Parametrization of the mixed bracket} 

For an arbitrary real $\epsilon$ the linear combination 
$\{\cdot,\cdot\}_1+\epsilon\{\cdot,\cdot\}_2$ allows the following
canonical parametrization (a 1-parameter deformation of (\ref{TL l par})):
\begin{equation}\label{TL mix par}
a_k=\eto{x_{k+1}\nm x_k\np\epsilon p_k},\qquad 
b_k=\frac{\eto{\epsilon p_k}-1}{\epsilon}+\epsilon\eto{x_k\nm x_{k-1}}
\end{equation}

\subsection{The flow TL}

Hamilton function:
\begin{equation}\label{H mix in xp}
H(x,p)=\epsilon^{-1}H_2(a,b)-\epsilon^{-1}\sum_kp_k=
\sum_{k} \frac{\eto{\epsilon p_k}-1-\epsilon p_k}{\epsilon^2}
+\sum_{k}\eto{x_k\nm x_{k-1}}
\end{equation}
Hamiltonian equations of motion:
\begin{equation}
\left\{\begin{array}{l}
\dot{x}_k = \Big(\eto{\epsilon p_k}-1\Big)/\epsilon\\ \\
\dot{p}_k = \eto{x_{k+1}\nm x_k}-\eto{x_k\nm x_{k-1}}
\end{array}\right.
\end{equation}
Newtonian equations of motion:
\begin{equation}
\ddot{x}_k=(1+\epsilon\dot{x}_k)\,\Big(\eto{x_{k+1}\nm x_k}-
\eto{x_k\nm x_{k-1}}\Big)
\end{equation}

\subsection{The map dTL}

Lagrangian form of equations of motion:
\begin{equation}
\left\{\begin{array}{l}
h\eto{\epsilon p_k} =\bigg(\epsilon\left(\eto{\wx_k\nm x_k}-1\right)+h\bigg)\,
\bigg(1+h\epsilon\eto{x_k\nm\wx_{k-1}}\bigg)\\ \\
h\eto{\epsilon\wip_k} = \bigg(\epsilon\left(\eto{\wx_k\nm x_k}-1\right)+h\bigg)\,
\bigg(1+h\epsilon\eto{x_{k+1}\nm\wx_k}\bigg)
\end{array}\right.
\end{equation}
Newtonian form of equations of motion:
\begin{equation}
\frac{\epsilon\Big(\eto{\wx_k\nm x_k}-1\Big)+h}
{\epsilon\Big(e\ueto{x_k\nm\undertilde{x_k}}-1\Big)+h}=
\frac{\bigg(1+h\epsilon e\ueto{\,\undertilde{x_{k+1}}\nm x_k}\bigg)}
{\bigg(1+h\epsilon\eto{x_k\nm \wx_{k-1}}\bigg)}
\end{equation}

\paragraph*{A special case $\epsilon=h$.}
If the parameter $\epsilon$ coincides with the (small) stepsize $h$, then the
previous discrete time system simplifies and serves as a discretization of the 
Toda lattice. The corresponding Lagrangian equations of motion:
\begin{equation}
\left\{\begin{array}{l}
\eto{hp_k} =\eto{\wx_k\nm x_k}\,\Big(1+h^2\eto{x_k\nm \wx_{k-1}}\Big)\\ \\
\eto{h\wip_k} = \eto{\wx_k\nm x_k}\,\Big(1+h^2\eto{x_{k+1}\nm \wx_k}\Big)
\end{array}\right.
\end{equation}
Newtonian form of equations of motion:
\begin{equation}
e\ueto{\wx_k\nm 2x_k\np\undertilde{x_k}}=
\frac{\bigg(1+h^2e\ueto{\,\undertilde{x_{k+1}}\nm x_k}\bigg)}
{\bigg(1+h^2\eto{x_k\nm\wx_{k-1}}\bigg)}
\end{equation}

\subsection{Lax representations and $r$--matrix structure}
The both systems above have Lax representations with the Lax matrix
\begin{equation}\label{mixed Toda L}
L_k=\left(\begin{array}{cc}
\lambda\eto{\epsilon p_k}-\lambda^{-1} & \epsilon\eto{x_k}\\ \\
-\epsilon\eto{-x_k} & \lambda\epsilon^2\end{array}\right)
\end{equation}
which satisfies the following $r$--matrix ansatz:
\begin{equation}\label{rmat eps ans}
\{L_k(\lambda)\stackrel{\otimes}{,}L_j(\mu)\}=\epsilon\Big[r(\lambda,\mu)\,,\,
L_k(\lambda)\otimes L_k(\mu)\Big]\delta_{kj}
\end{equation}
with the $r$--matrix (\ref{r}).

The Lax representation for the flow TL: (\ref{Lax}) with
\begin{equation}
M_k=\left(\begin{array}{cc}
\displaystyle\frac{\lambda^{-2}-1}{\epsilon}+\epsilon\eto{x_k-x_{k-1}} & 
-\lambda^{-1}\eto{x_k}\\ \\ 
\lambda^{-1}\eto{-x_{k-1}} & 0\end{array}\right)
\end{equation}
The Lax representation for the map dTL: (\ref{dLax}) with
\begin{equation}
V_k=\left(\begin{array}{cc}
1-\displaystyle\frac{h}{\epsilon}+\displaystyle\frac{h}{\epsilon}\lambda^{-2}
+h(\epsilon-h)\eto{x_k\nm\wx_{k-1}}
 & -h\lambda^{-1}\eto{x_k}\\ \\
h\lambda^{-1}\eto{-\wx_{k-1}} & 1\end{array}\right)
\end{equation}

\paragraph*{Special case $\epsilon =h$.} In this case also the Lax 
representation of the map dTL simplifies significantly: it reads (\ref{dLax})
with
\begin{equation}
L_k=\left(\begin{array}{cc}
\lambda\eto{hp_k}-\lambda^{-1} & h\eto{x_k}\\ \\
-h\eto{-x_k} & \lambda h^2\end{array}\right)
\end{equation}
\begin{equation}
V_k=\left(\begin{array}{cc}
\lambda^{-2} & -h\lambda^{-1}\eto{x_k}\\ \\
h\lambda^{-1}\eto{-\wx_{k-1}} & 1\end{array}\right)
\end{equation}

\setcounter{equation}{0}
\section{Explicit dRTL: different parametrizations}

\subsection{Hirota's discretization of the Toda lattice}
This system corresponds to the following parametrization of the $(c,d)$
variables of the relativistic Toda hierarchy:
\begin{equation}\label{eRTL l par}
c_k=h^2\eto{x_{k+1}\nm x_k},\quad d_k=1+hp_k-h^2\eto{x_k\nm x_{k-1}}
\end{equation}
This results in the Poisson bracket $h\{\cdot,\cdot\}_1$.

The equations of motion of the explicit dRTL map in this parametrization may be 
put in the following Lagrangian form:
\begin{equation}
\left\{\begin{array}{l}
hp_k = \eto{\wx_k\nm x_k}-1-h^2\eto{x_{k+1}\nm x_k}+h^2\eto{x_k\nm x_{k-1}}\\ \\
h\wip_k = \eto{\wx_k\nm x_k}-1
\end{array}\right.
\end{equation}
It follows a nice ''Hamiltonian'' form of equations of motion:
\begin{equation}
\left\{\begin{array}{l}
\eto{\wx_k\nm x_k}-1 = h\wip_k\\ \\
\wip_k-p_k = h\eto{x_{k+1}\nm x_k}-h\eto{x_k\nm x_{k-1}}
\end{array}\right.
\end{equation}
The Newtonian form of equations of motion:
\begin{equation}\label{Hir}
e\ueto{\wx_k\nm x_k}-e\ueto{x_k-\undertilde{x_k}}=
h^2\bigg(e\ueto{x_{k+1}\nm x_k}-e\ueto{x_k-x_{k-1}}\bigg)
\end{equation}
The Lax representation: (\ref{dLax}) with
\begin{equation}\label{Hir L}
L_k=\left(\begin{array}{cc}
p_k+\lambda & \eto{x_k}\\ \\-(1+hp_k)\eto{-x_k} & -h\end{array}\right)
\end{equation}
\begin{equation}
V_k=\left(\begin{array}{cc}
1-h\lambda-h^2\eto{x_k\nm x_{k-1}} & -h\eto{x_k}\\ \\
h\eto{-x_{k-1}} & 1\end{array}\right)
\end{equation}
Note that the matrix $V_k$ depends only on the $x_j$'s, and not on their
discrete time updates $\wx_j$ or on the momenta $p_j$. This will be the 
common feature of all the results in this Section.

The Lax matrix (\ref{Hir L}) is a one--parameter deformation of the standard
Toda Lax matrix (\ref{Toda L}), but still satisfies the $r$--matrix ansatz 
(\ref{rmat ans 1})

\subsection{Standard discretization of the Toda lattice}

This system corresponds to the following parametrization of the variables 
$(c,d)$ of the relativistic Toda hierarchy:
\begin{equation}\label{r q par}
c_k=h^2\eto{x_{k+1}\nm x_k\np hp_k}, \qquad d_k=\eto{hp_k}
\end{equation}
This results in the Poisson bracket $h\{\cdot,\cdot\}_2$. The equations of 
motion of the explicit dRTL map may be presented in this parametrization in the
following Lagrange form:
\begin{equation}
\left\{\begin{array}{l}
\eto{hp_k} =\eto{\wx_k\nm x_k}\,
\displaystyle\frac{\Big(1+h^2\eto{x_k\nm x_{k-1}}\Big)}
{\Big(1+h^2\eto{x_{k+1}\nm x_k}\Big)}\\ \\
\eto{h\wip_k} =\eto{\wx_k\nm x_k}
\end{array}\right.
\end{equation}
This may be presented in a nice ''Hamiltonian'' form:
\begin{equation}
\left\{\begin{array}{l}
\wx_k-x_k=h\wip_k\\ \\
\eto{h\wip_k\nm hp_k}=
\displaystyle\frac{\Big(1+h^2\eto{x_{k+1}\nm x_k}\Big)}
{\Big(1+h^2\eto{x_k\nm x_{k-1}}\Big)}
\end{array}\right.
\end{equation}
The Newtonian form of equations of motion:
\begin{equation}\label{stand}
e\ueto{\wx_k\nm 2x_k\np\undertilde{x_k}}=
\frac{\Big(1+h^2\eto{x_{k+1}\nm x_k}\Big)}
{\Big(1+h^2\eto{x_k\nm x_{k-1}}\Big)}
\end{equation}
The Lax representation: (\ref{dLax}) with
\begin{equation}\label{stand L}
L_k=\left(\begin{array}{cc}
\lambda\eto{hp_k}-\lambda^{-1} & h\eto{x_k}\\ \\
-h\eto{-x_k\np hp_k} & 0\end{array}\right)
\end{equation}
\begin{equation}
V_k=\left(\begin{array}{cc}
\lambda^{-2} & -h\lambda^{-1}\eto{x_k}\\ \\
h\lambda^{-1}\eto{-x_{k-1}} & 1\end{array}\right)
\end{equation}
The Lax matrix (\ref{stand L}) is also a 1--parameter deformation of the
standard Toda Lax matrix (\ref{Toda L}), but satisfies a different 
$r$--matrix ansatz:
\begin{equation}
\{L_k(\lambda)\stackrel{\otimes}{,}L_j(\mu)\}=h\Big[r(\lambda,\mu)\,,\,
L_k(\lambda)\otimes L_k(\mu)\Big]\delta_{kj}
\end{equation}
with the $r$--matrix (\ref{r}).

\subsection{Explicit discretization of the modified Toda lattice}

This system corresponds to the following parametrization of the variables 
$(c,d)$ of the relativistic Toda hierarchy:
\begin{equation}\label{emToda par}
c_k=h^2\eto{x_{k+1}\nm x_k\np p_k},\qquad 
d_k=1+h\eto{p_k}+h\eto{x_k\nm x_{k-1}}
\end{equation}
This results in the Poisson bracket $\{\cdot,\cdot\}_2-\{\cdot,\cdot\}_1$. 
The equations of motion of the explicit dRTL map may be presented in this 
parametrization in the following Lagrange form:
\begin{equation}
\left\{\begin{array}{l}
h\eto{p_k} =\Big(\eto{\wx_k\nm x_k}-1\Big)\,
\displaystyle\frac{\Big(1+h\eto{x_k\nm x_{k-1}}\Big)}
{\Big(1+h\eto{x_{k+1}\nm x_k}\Big)}\\ \\
h\eto{\wip_k} =\Big(\eto{\wx_k\nm x_k}-1\Big)
\end{array}\right.
\end{equation}
This may be presented in a nice ''Hamiltonian'' form:
\begin{equation}
\left\{\begin{array}{l}
\eto{\wx_k\nm x_k}-1=h\eto{\wip_k}\\ \\
\eto{\wip_k\nm p_k}=
\displaystyle\frac{\Big(1+h\eto{x_{k+1}\nm x_k}\Big)}
{\Big(1+h\eto{x_k\nm x_{k-1}}\Big)}
\end{array}\right.
\end{equation}
The Newtonian form of equations of motion:
\begin{equation}\label{emToda}
\frac{\Big(e\ueto{\wx_k\nm x_k}-1\Big)}
{\Big(e\ueto{x_k\nm\undertilde{x_k}}-1\Big)}
=\frac{\Big(1+h\eto{x_{k+1}\nm x_k}\Big)}{\Big(1+h\eto{x_k\nm x_{k-1}}\Big)}
\end{equation}
The Lax representation: (\ref{dLax}) with
\begin{equation}\label{emToda L}
L_k=\left(\begin{array}{cc}
\lambda\eto{p_k}-\lambda^{-1} & \eto{x_k}\\ \\
-\Big(1+h\eto{p_k}\Big)\eto{-x_k} & \lambda\end{array}\right)
\end{equation}
\begin{equation}
V_k=\left(\begin{array}{cc}
1+h\lambda^{-2}+h\eto{x_k\nm x_{k-1}} & -h\lambda^{-1}\eto{x_k}\\ \\
h\lambda^{-1}\eto{-x_{k-1}} & 1\end{array}\right)
\end{equation}
The Lax matrix (\ref{emToda L}) is a 1--parameter deformation of the
Lax matrix (\ref{mToda L}), but satisfies the same $r$--matrix ansatz
(\ref{rmat 2 ans}).

\setcounter{equation}{0}
\section{Systems RTL$\pm$ and dRTL$\pm$: \newline
Parametrization of the quadratic bracket}

Canonical parametrization of the bracket $\{\cdot,\cdot\}_2$:
\begin{equation}\label{RTL q par}
d_k=\eto{p_k}, \qquad c_k=g^2\eto{x_{k+1}\nm x_k\np p_k}
\end{equation}
The resulting system is the {\it relativistic Toda lattice}.

\subsection{The flow RTL+}
Hamilton function:
\begin{equation}\label{RTL H2+ in xp}
H(x,p)=H_2^{(+)}(c,d)=\sum_{k} \eto{p_k}\Big(1+g^2\eto{x_{k+1}\nm x_k}\Big)
\end{equation}
Hamiltonian equations of motion:
\begin{equation}\label{RTL+ q Ham}
\left\{\begin{array}{l}
\dot{x}_k=\eto{p_k}\Big(1+g^2\eto{x_{k+1}\nm x_k}\Big)\\ \\
\dot{p}_k=g^2\eto{x_{k+1}\nm x_k\np p_k}-g^2\eto{x_k\nm x_{k-1}\np p_{k-1}}
\end{array}\right.
\end{equation}
Lagrangian equations of motion:
\begin{equation}\label{RTL+ q Lagr}
\left\{\begin{array}{l}
\eto{p_k}=\displaystyle\frac{\dot{x}_k}
{\Big(1+g^2\eto{x_{k+1}\nm x_k}\Big)}\\ \\
\dot{p}_k=\dot{x}_{k}\,\displaystyle\frac{g^2\eto{x_{k+1}\nm x_k}}
{\Big(1+g^2\eto{x_{k+1}\nm x_k}\Big)}-\dot{x}_{k-1}\,\displaystyle\frac
{g^2\eto{x_k\nm x_{k-1}}}{\Big(1+g^2\eto{x_k\nm x_{k-1}}\Big)}
\end{array}\right.
\end{equation}
Newtonian equations of motion:
\begin{equation}\label{RTL New}
\ddot{x}_k=
\dot{x}_{k+1}\dot{x}_k\,\frac{g^2\eto{x_{k+1}\nm x_k}}
{\Big(1+g^2\eto{x_{k+1}\nm x_k}\Big)}-
\dot{x}_k\dot{x}_{k-1}\,\frac{g^2\eto{x_k\nm x_{k-1}}}
{\Big(1+g^2\eto{x_k\nm x_{k-1}}\Big)}
\end{equation}

\subsection{The map dRTL+}

''Hamiltonian'' form of equations of motion:
\begin{equation}
\left\{\begin{array}{l}
\eto{\wx_k\nm x_k}-1=h\eto{\wip_k}\Big(1+g^2\eto{x_{k+1}\nm\wx_k}\Big)\\ \\
\eto{\wip_k-p_k} = 
\displaystyle\frac{\bigg(1+hg^2\eto{x_{k+1}\nm \wx_k\np \wip_k}\bigg)}
{\bigg(1+hg^2\eto{x_k\nm \wx_{k-1}\np\wip_{k-1}}\bigg)}
\end{array}\right.
\end{equation}
Lagrangian form of equations of motion:
\begin{equation}
\left\{\begin{array}{l}
h\eto{p_k} = 
\displaystyle\frac{\bigg(\eto{\wx_k\nm x_k}-1\bigg)}
{\bigg(1+g^2\eto{x_k\nm \wx_{k-1}}\bigg)}\:
\frac{\bigg(1+g^2\eto{x_k\nm x_{k-1}}\bigg)}
{\bigg(1+g^2\eto{x_{k+1}\nm x_k}\bigg)}\\ \\
h\eto{\wip_k} =
\displaystyle\frac{\bigg(\eto{\wx_k\nm x_k}-1\bigg)}
{\bigg(1+g^2\eto{x_{k+1}\nm\wx_k}\bigg)}
\end{array}\right.
\end{equation}
Newtonian form of equations of motion:
\begin{equation}\label{dRTL New}
\displaystyle
\frac{\bigg(\eto{\wx_k\nm x_k}-1\bigg)}
{\bigg(e\ueto{x_k\nm \undertilde{x_k}}-1\bigg)}=
\displaystyle\frac{\bigg(1+g^2\eto{x_{k+1}\nm x_k}\bigg)\,
\bigg(1+g^2\eto{x_k\nm\wx_{k-1}}\bigg)}
{\bigg(1+g^2e\ueto{\,\undertilde{x_{k+1}}\nm x_k}\bigg)\,
\bigg(1+g^2e\ueto{x_k\nm x_{k-1}}\bigg)}
\end{equation}

\subsection{The flow RTL$-$}
Hamilton function:
\begin{equation}\label{RTL H2- in xp}
H(x,p)=H_2^{(-)}(c,d)=\sum_{k} \eto{-p_k}\Big(1+g^2\eto{x_{k}\nm x_{k-1}}\Big)
\end{equation}
Hamiltonian equations of motion:
\begin{equation}\label{RTL- q Ham}
\left\{\begin{array}{l}
\dot{x}_k=-\eto{-p_k}\Big(1+g^2\eto{x_{k}\nm x_{k-1}}\Big)\\ \\
\dot{p}_k=g^2\eto{x_{k+1}\nm x_k\nm p_{k+1}}-g^2\eto{x_k\nm x_{k-1}\nm p_{k}}
\end{array}\right.
\end{equation}
Lagrangian equations of motion:
\begin{equation}\label{RTL- q Lagr}
\left\{\begin{array}{l}
\eto{p_k}=-\displaystyle\frac{\Big(1+g^2\eto{x_{k}\nm x_{k-1}}\Big)}
{\dot{x}_k}\\ \\
\dot{p}_k=-\dot{x}_{k+1}\,\displaystyle\frac{g^2\eto{x_{k+1}\nm x_k}}
{\Big(1+g^2\eto{x_{k+1}\nm x_k}\Big)}+\dot{x}_k\,\displaystyle\frac
{g^2\eto{x_k\nm x_{k-1}}}{\Big(1+g^2\eto{x_k\nm x_{k-1}}\Big)}
\end{array}\right.
\end{equation}
Newtonian equations of motion -- the same (\ref{RTL New}) as for the RTL+ (!).

\subsection{The map dRTL$-$}
''Hamiltonian'' form of equations of motion:
\begin{equation}
\left\{\begin{array}{l}
\eto{\wx_k\nm x_k}-1=-h\eto{-p_k}\Big(1+g^2\eto{x_{k}\nm\wx_{k-1}}\Big)\\ \\
\eto{\wip_k-p_k} = 
\displaystyle\frac{\bigg(1-hg^2\eto{x_k\nm \wx_{k-1}\nm p_k}\bigg)}
{\bigg(1-hg^2\eto{x_{k+1}\nm \wx_k\nm p_{k+1}}\bigg)}
\end{array}\right.
\end{equation}
Lagrangian form of equations of motion:
\begin{equation}
\left\{\begin{array}{l}
\eto{p_k}= -\,\displaystyle\frac{h\bigg(1+g^2\eto{x_k\nm\wx_{k-1}}\bigg)}
{\bigg(\eto{\wx_k\nm x_k}-1\bigg)}\\ \\
\eto{\wip_k}= -\,\displaystyle\frac{h\bigg(1+g^2\eto{x_{k+1}\nm \wx_k}\bigg)}
{\bigg(\eto{\wx_k\nm x_k}-1\bigg)}\:
\displaystyle\frac{\bigg(1+g^2\eto{\wx_k\nm \wx_{k-1}}\bigg)}
{\bigg(1+g^2\eto{\wx_{k+1}\nm \wx_k}\bigg)}
\end{array}\right.
\end{equation}
Newtonian form of equations of motion  -- (\ref{dRTL New}), the same as for the 
dRTL+ (!).

\subsection{Lax representations and $r$--matrix structure}
Lax representations for the four systems of this Section is given in terms
one and the same Lax matrix:
\begin{equation}\label{rToda L}
L_k=\left(\begin{array}{cc}
\lambda\eto{p_k}-\lambda^{-1} & \eto{x_k}\\ \\
-g^2\eto{-x_k\np p_k} & 0\end{array}\right)
\end{equation}
It satisfies the $r$--matrix ansatz (\ref{rmat 2 ans}) with the $r$--matrix 
(\ref{r}).

Lax representation for the flow RTL+: (\ref{Lax}) with
\begin{equation}\label{rToda+ M}
M_k=\left(\begin{array}{cc}
\lambda^{-2}+g^2\eto{x_k\nm x_{k-1}\np p_{k-1}} & -\lambda^{-1}\eto{x_k}\\ \\ 
\lambda^{-1}g^2\eto{-x_{k-1}\np p_{k-1}} & 0\end{array}\right)
\end{equation}
Lax representation for the map dRTL+: (\ref{dLax}) with
\begin{equation}
V_k=\left(\begin{array}{cc}
1+h\lambda^{-2}+hg^2\eto{x_k\nm\wx_{k-1}\np\wip_{k-1}} & 
-h\lambda^{-1}\eto{x_k}\\ \\ 
h\lambda^{-1}g^2\eto{-\wx_{k-1}\np\wip_{k-1}} & 1\end{array}\right)
\end{equation}
Lax representation for the flow RTL$-$: (\ref{Lax}) with 
\begin{equation}\label{rToda- M}
M_k=\left(\begin{array}{cc}
0 & -\lambda\eto{x_k\nm p_k}\\ \\ 
\lambda g^2\eto{-x_{k-1}} & \lambda^2+g^2\eto{x_k\nm x_{k-1}\nm p_{k}}
\end{array}\right)
\end{equation}
Lax representation for the map dRTL$-$: 
\begin{equation}\label{dLax inv}
W_{k+1}\widetilde{L}_k=L_kW_k
\end{equation} 
with the same Lax matrix (\ref{rToda L}) and
\begin{equation}
W_k=\left(\begin{array}{cc}
1 & h\lambda\eto{x_k\nm p_k}\\ \\
-h\lambda g^2\eto{-\wx_{k-1}} & 
1-h\lambda^{2}-hg^2\eto{x_k\nm \wx_{k-1}\nm p_k} \end{array}\right)
\end{equation}

\setcounter{equation}{0}
\section{Systems RTL$\pm$ and dRTL$\pm$: \newline
Parametrization of the linear bracket}

Canonical parametrization of the bracket $\{\cdot,\cdot\}_1$:
\begin{equation}\label{RTL l par}
d_k=p_k-\eto{x_k\nm x_{k-1}},\qquad c_k=\eto{x_{k+1}\nm x_k}
\end{equation}

\subsection{The flow RTL+}
Hamilton function:
\begin{equation}\label{RTL H1+ in xp}
H(x,p)=H_1^{(+)}(c,d)=\frac{1}{2}\sum_{k}p_k^2+\sum_{k}p_k\eto{x_{k+1}\nm x_k}
\end{equation}
Hamiltonian equations of motion:
\begin{equation}\label{RTL+ l Ham}
\left\{\begin{array}{l}
\dot{x}_k = p_k+\eto{x_{k+1}\nm x_k}\\ \\
\dot{p}_k = p_k\eto{x_{k+1}\nm x_k}-p_{k-1}\eto{x_k\nm x_{k-1}}
\end{array}\right.
\end{equation}
Lagrangian equations of motion:
\begin{equation}\label{RTL+ l Lagr}
\left\{\begin{array}{l}
p_k=\dot{x}_k-\eto{x_{k+1}\nm x_k}\\ \\
\dot{p}_k =\left(\dot{x}_{k}-\eto{x_{k+1}\nm x_k}\right)\eto{x_{k+1}\nm x_k}
-\left(\dot{x}_{k-1}-\eto{x_k\nm x_{k-1}}\right)\eto{x_k\nm x_{k-1}} 
\end{array}\right.
\end{equation}
Newtonian equations of motion:
\begin{equation}\label{RTL+ l New}
\ddot{x}_k = \left(\dot{x}_{k+1}-\eto{x_{k+1}\nm x_k}\right)
\eto{x_{k+1}\nm x_k}
-\left(\dot{x}_{k-1}-\eto{x_k\nm x_{k-1}}\right)\eto{x_k\nm x_{k-1}}
\end{equation}

\subsection{The map dRTL+}

''Hamiltonian'' form of equations of motion:
\begin{equation}
\left\{\begin{array}{l}
\eto{\wx_k\nm x_k}-1=h\wip_k +
\displaystyle\frac{h\eto{x_{k+1}\nm\wx_k}}
{\bigg(1-h\eto{x_{k+1}\nm\wx_k}\bigg)}\\ \\
\wip_k-p_k = h\wip_k\eto{x_{k+1}\nm\wx_k}-h\wip_{k-1}\eto{x_k\nm\wx_{k-1}}
\end{array}\right.
\end{equation}
Lagrangian form of equations of motion:
\begin{equation}
\left\{\begin{array}{l}
hp_k = \eto{\wx_k\nm x_k}-
\displaystyle\frac{1}{\bigg(1-h\eto{x_k\nm \wx_{k-1}}\bigg)}
     +h\eto{x_k\nm x_{k-1}}-h\eto{x_{k+1}\nm x_k}\\ \\
h\wip_k = \eto{\wx_k\nm x_k}-
\displaystyle\frac{1}{\bigg(1-h\eto{x_{k+1}\nm \wx_k}\bigg)}
\end{array}\right.
\end{equation}
Newtonian form of equations of motion:
\begin{equation}
e\ueto{\wx_k\nm x_k}-e\ueto{x_k\nm\undertilde{x_k}}=
he\ueto{x_{k+1}\nm x_k}-
\displaystyle\frac{he\ueto{\,\undertilde{x_{k+1}}\nm x_k}}
{\bigg(1-he\ueto{\,\undertilde{x_{k+1}}\nm x_k}\bigg)}
-h\eto{x_k\nm x_{k-1}}+\displaystyle\frac{h\eto{x_k\nm \wx_{k-1}}}
{\bigg(1-h\eto{x_k\nm \wx_{k-1}}\bigg)}
\label{RTL dnew1}
\end{equation}

\subsection{The flow RTL$-$}

Hamilton function:
\begin{equation}\label{RTL H1- in xp}
H(x,p)=H_1^{(-)}(c,d)=-\sum_{k}\log\Big(p_k-\eto{x_k\nm x_{k-1}}\Big)
\end{equation}
Hamiltonian equations of motion:
\begin{equation}\label{RTL- l Ham}
\left\{\begin{array}{l}
\dot{x}_k = -\displaystyle\frac{1}{\Big(p_k-\eto{x_k\nm x_{k-1}}\Big)}\\ \\
\dot{p}_k = 
\displaystyle\frac{\eto{x_{k+1}\nm x_k}}{\Big(p_{k+1}-\eto{x_{k+1}\nm x_k}\Big)}
-\displaystyle\frac{\eto{x_k\nm x_{k-1}}}{\Big(p_k-\eto{x_k\nm x_{k-1}}\Big)}
\end{array}\right.
\end{equation}
Lagrangian equations of motion:
\begin{equation}\label{RTL- l Lagr}
\left\{\begin{array}{l}
p_k=-\displaystyle\frac{1}{\dot{x}_k}+\eto{x_k\nm x_{k-1}} \\ \\
\dot{p}_k = -\dot{x}_{k+1}\eto{x_{k+1}\nm x_k}+
\dot{x}_{k}\eto{x_k\nm x_{k-1}}
\end{array}\right.
\end{equation}
Newtonian form of equations of motion:
\begin{equation}\label{RTL- l New}
\ddot{x}_k=-\dot{x}_k^2\Big(\dot{x}_{k+1}\eto{x_{k+1}\nm x_k}-
\dot{x}_{k-1}\eto{x_k\nm x_{k-1}}\Big)
\end{equation}

\subsection{The map dRTL$-$}

''Hamiltonian'' form of equations of motion:
\begin{equation}
\left\{\begin{array}{l}
\eto{\wx_k\nm x_k}-1= -\,\displaystyle\frac{h}
{\bigg(p_k-\eto{x_k\nm\wx_{k-1}}\bigg)}\\ \\
\wip_k-p_k  = 
\displaystyle\frac{h\eto{x_{k+1}\nm\wx_k}}
{\bigg(p_{k+1}-\eto{x_{k+1}\nm\wx_k}\bigg)}-
\displaystyle\frac{h\eto{x_k\nm\wx_{k-1}}}
{\bigg(p_k-\eto{x_k\nm\wx_{k-1}}\bigg)}
\end{array}\right.
\end{equation}
Lagrangian form of equations of motion:
\begin{equation}
\left\{\begin{array}{l}
p_k = -\displaystyle\frac{h}{\bigg(\eto{\wx_k\nm x_k}-1\bigg)}
+\eto{x_k\nm\wx_{k-1}}\\ \\
\wip_k  = -\displaystyle\frac{h}{\bigg(\eto{\wx_k\nm x_k}-1\bigg)}
+\eto{x_{k+1}\nm\wx_k}-\eto{\wx_{k+1}\nm\wx_k}+\eto{\wx_k\nm\wx_{k-1}}
\end{array}\right.
\end{equation}
Newtonian form of equations of motion:
\begin{equation}\label{dRTL- l New}
\frac{h}{\bigg(\eto{\wx_k\nm x_k}-1\bigg)}\,-\,
\frac{h}{\bigg(e\ueto{x_k\nm\undertilde{x_k}}-1\bigg)}=
e\ueto{x_{k+1}\nm x_k}-e\ueto{\,\undertilde{x_{k+1}}\nm x_k}-
e\ueto{x_k\nm x_{k-1}}+e\ueto{x_k\nm \wx_{k-1}}
\end{equation}

\subsection{Lax representations and $r$--matrix structure}
All four systems considered in this Section allow Lax representations with
the Lax matrix
\begin{equation}\label{RTL l L}
L_k=\left(\begin{array}{cc}
p_k+\lambda & \eto{x_k} \\ \\ -p_k\eto{-x_k} & -1
\end{array}\right)
\end{equation}
which satisfies the $r$--matrix ansatz (\ref{rmat ans 1}).

Lax representation for the flow RTL+: (\ref{Lax}) with
\begin{equation}\label{RTL+ l M}
M_k=\left(\begin{array}{cc}
-\lambda & -\eto{x_k} \\ \\ p_{k-1}\eto{-x_{k-1}} & 0
\end{array}\right)
\end{equation}
Lax representation for the map dRTL+: (\ref{dLax}) with 
\begin{equation}
V_k=\left(\begin{array}{cc}
1-h\lambda-h^2\wip_{k-1}\eto{x_k\nm\wx_{k-1}} & -h\eto{x_k} \\ \\ 
h\wip_{k-1}\eto{-\wx_{k-1}} & 1
\end{array}\right)
\end{equation}
Lax representation for the flow RTL$-$: (\ref{Lax}) with 
\begin{equation}\label{RTL- l M}
M_k=\frac{\lambda^{-1}}{\Big(p_k-\eto{x_k\nm x_{k-1}}\Big)}
\left(\begin{array}{cc}
p_k & \eto{x_k} \\ \\ -p_k\eto{-x_{k-1}} & -\eto{x_k\nm x_{k-1}}
\end{array}\right)
\end{equation}
Lax representation the map dRTL$-$: (\ref{dLax inv}) with 
\begin{equation}
W_k=I-\frac{h}{\lambda+h}\,\frac{1}{\bigg(p_k-\eto{x_k\nm\wx_{k-1}}\bigg)}
\left(\begin{array}{cc}
p_k & \eto{x_k} \\ \\ -p_k\eto{-\wx_{k-1}} & -\eto{x_k\nm\wx_{k-1}}
\end{array}\right)
\end{equation}
This last Lax representation may be also presented in the form (\ref{dLax})
with matrix $V_k$ being more nice than $W_k$ (which is non-tipical for the
map dRTL$-$):
\begin{equation}
V_k=I+\displaystyle\frac{h\lambda^{-1}}{\bigg(p_k-\eto{x_k\nm\wx_{k-1}}\bigg)}
\left(\begin{array}{cc}
p_k & \eto{x_k} \\ \\ -p_k\eto{-\wx_{k-1}} & -\eto{x_k\nm\wx_{k-1}}
\end{array}\right)
\end{equation}

\setcounter{equation}{0}
\section{Systems RTL$\pm$ and dRTL$\pm$:\newline
The first mixed parametrization} 

We now consider the following canonical parametrization of the bracket
$\{\cdot,\cdot\}_2-\delta\{\cdot,\cdot\}_1$:
\begin{equation}\label{RTL gq par}
d_k=\eto{p_k}+\delta\Big(1+g^2\eto{x_k\nm x_{k-1}}\Big) \qquad
c_k=g^2\eto{x_{k+1}\nm x_k\np p_k}
\end{equation}
This is, obviously, a 1--parameter deformation of (\ref{RTL q par}).
The results of this section in the limit $\delta\to 0$ imply the results
of the Section 8.

\subsection{The flow RTL+}
Hamilton function:
\begin{equation}
H(x,p)=H_2^{(+)}(c,d)=
\sum_{k}\eto{p_k}\Big(1+g^2\eto{x_{k+1}\nm x_k}\Big)+
\delta g^2\sum_{k}\eto{x_k\nm x_{k-1}}
\end{equation}
Hamiltonian equations of motion:
\begin{equation}\label{RTL+ gq Ham}
\left\{\begin{array}{l}
\dot{x}_k=\eto{p_k}\Big(1+g^2\eto{x_{k+1}\nm x_k}\Big)\\ \\
\dot{p}_k=g^2\Big(\eto{p_k}+\delta\Big)\eto{x_{k+1}\nm x_k}-
g^2\Big(\eto{p_{k-1}}+\delta\Big)\eto{x_k\nm x_{k-1}}
\end{array}\right.
\end{equation}
Lagrangian equations of motion:
\begin{equation}\label{RTL+ gq Lagr}
\left\{\begin{array}{l}
\eto{p_k}=\displaystyle\frac{\dot{x}_k}{\Big(1+g^2\eto{x_{k+1}\nm x_k}\Big)}\\ \\
\dot{p}_k=\dot{x}_k\,\displaystyle\frac{g^2\eto{x_{k+1}\nm x_k}}
{\Big(1+g^2\eto{x_{k+1}\nm x_k}\Big)}-\dot{x}_{k-1}\,\displaystyle\frac
{g^2\eto{x_k\nm x_{k-1}}}{\Big(1+g^2\eto{x_k\nm x_{k-1}}\Big)}
+\delta g^2\Big(\eto{x_{k+1}\nm x_k}-\eto{x_k\nm x_{k-1}}\Big)
\end{array}\right.
\end{equation}
Newtonian equations of motion:
\begin{eqnarray}
\ddot{x}_k & = &
\dot{x}_{k+1}\dot{x}_k\,\frac{g^2\eto{x_{k+1}\nm x_k}}
{\Big(1+g^2\eto{x_{k+1}\nm x_k}\Big)}
-\dot{x}_k\dot{x}_{k-1}\,\frac{g^2\eto{x_{k}\nm x_{k-1}}}
{\Big(1+g^2\eto{x_{k}\nm x_{k-1}}\Big)}\nonumber\\
&&+\delta g^2\dot{x}_k\Big(\eto{x_{k+1}\nm x_k}-\eto{x_k\nm x_{k-1}}\Big)
\label{RTL+ gq New}
\end{eqnarray}
It is interesting to remark that the right--hand side of this equation 
coincides with the right--hand side of (\ref{RTL New}) plus an {\it additive} 
perturbation which is exactly the right--hand side of the modified Toda
lattice (\ref{mToda New}).

\subsection{The map dRTL+}

''Hamiltonian'' form of equations of motion:
\begin{equation}
\left\{\begin{array}{l}
\eto{\wx_k\nm x_k}-1=
\displaystyle\frac{h\eto{\wip_k}\Big(1+g^2\eto{x_{k+1}\nm\wx_k}\Big)}
{\bigg(1+h\delta g^2\eto{x_{k+1}\nm \wx_{k}}\bigg)}\\ \\
\eto{\wip_k\nm p_k} = 
\displaystyle\frac
{\bigg(1+hg^2\Big(\eto{\wip_k}+\delta\Big)\eto{x_{k+1}\nm\wx_k}\bigg)}
{\bigg(1+hg^2\Big(\eto{\wip_{k-1}}+\delta\Big)\eto{x_k\nm\wx_{k-1}}\bigg)}
\end{array}\right.
\end{equation}
Lagrangian form of equations of motion:
\begin{equation}
\left\{\begin{array}{l}
h\eto{p_k} = 
\displaystyle\frac{\bigg(\eto{\wx_k\nm x_k}-1\bigg)}
{\bigg(1+g^2\eto{x_k\nm\wx_{k-1}}\bigg)}\:
\frac{\bigg(1+g^2\eto{x_k\nm x_{k-1}}\bigg)}
{\bigg(1+g^2\eto{x_{k+1}\nm x_k}\bigg)}\;
\bigg(1+h\delta g^2\eto{x_k\nm \wx_{k-1}}\bigg)\\ \\
h\eto{\wip_k} =
\displaystyle\frac{\bigg(\eto{\wx_k\nm x_k}-1\bigg)}
{\bigg(1+g^2\eto{x_{k+1}\nm\wx_k}\bigg)}\:
\bigg(1+h\delta g^2\eto{x_{k+1}\nm \wx_{k}}\bigg)
\end{array}\right.
\end{equation}
Newtonian form of equations of motion:
\begin{equation}\label{dRTL+ gq New}
\displaystyle\frac{\bigg(e\ueto{\wx_k\nm x_k}-1\bigg)}
{\bigg(e\ueto{x_k\nm\undertilde{x_k}}-1\bigg)} 
= \displaystyle\frac{\bigg(1+g^2\eto{x_{k+1}\nm x_k}\bigg)\,
\bigg(1+g^2\eto{x_k\nm\wx_{k-1}}\bigg)\,
\bigg(1+h\delta g^2e\ueto{\,\undertilde{x_{k+1}}\nm x_k}\bigg)}
{\bigg(1+g^2e\ueto{\,\undertilde{x_{k+1}}\nm x_k}\bigg)\,
\bigg(1+g^2\eto{x_k\nm x_{k-1}}\bigg)\,
\bigg(1+h\delta g^2\eto{x_{k}\nm\wx_{k-1}}\bigg)}
\end{equation}
Interesting enough, the right--hand side of the last equation is the {\it 
product} of the right--hand sides of (\ref{dRTL New}) and (\ref{dmToda New}).

\subsection{The flow RTL$-$}
Hamilton function:
\begin{equation}
H(x,p)=-\delta^{-1}H_1^{(-)}(c,d)-\delta^{-1}\sum_{k}p_k=
\delta^{-1}\sum_k \log(d_k)-\delta^{-1}\sum_k p_k
\end{equation}
where, recall,
\begin{equation}
d_k=\eto{p_k}+\delta\Big(1+g^2\eto{x_k\nm x_{k-1}}\Big)=\eto{p_k}+O(\delta)
\end{equation}
Hamiltonian equations of motion:
\begin{equation}\label{RTL- gq Ham}
\left\{\begin{array}{l}
\dot{x}_k=-d_k^{-1}\Big(1+g^2\eto{x_k\nm x_{k-1}}\Big)\\ \\
\dot{p}_k=d_{k+1}^{-1}g^2\eto{x_{k+1}\nm x_{k}}
-d_k^{-1}g^2\eto{x_k\nm x_{k-1}}
\end{array}\right.
\end{equation}
Lagrangian equations of motion:
\begin{equation}\label{RTL- gq Lagr}
\left\{\begin{array}{l}
\eto{p_k}=-\,\displaystyle\frac
{\Big(1+g^2\eto{x_k\nm x_{k-1}}\Big)}{\dot{x}_k}\,(1+\delta\dot{x}_k)\\ \\
\dot{p}_k=-\dot{x}_{k+1}\,\displaystyle\frac{g^2\eto{x_{k+1}\nm x_k}}
{\Big(1+g^2\eto{x_{k+1}\nm x_k}\Big)}+\dot{x}_k\,\displaystyle\frac
{g^2\eto{x_k\nm x_{k-1}}}{\Big(1+g^2\eto{x_k\nm x_{k-1}}\Big)}
\end{array}\right.
\end{equation}
Newtonian equations of motion:
\begin{equation}\label{RTL- gq New}
\ddot{x}_k =
(1+\delta\dot{x}_k)\left(
\dot{x}_{k+1}\dot{x}_k\,\frac{g^2\eto{x_{k+1}\nm x_k}}
{\Big(1+g^2\eto{x_{k+1}\nm x_k}\Big)}-
\dot{x}_k\dot{x}_{k-1}\,\frac{g^2\eto{x_k\nm x_{k-1}}}
{\Big(1+g^2\eto{x_k\nm x_{k-1}}\Big)}
\right)
\end{equation}
This is a {\it multiplicative} perturbation of (\ref{RTL New}).

\subsection{The map dRTL$-$}

''Hamiltonian'' equations of motion:
\begin{equation}
\left\{\begin{array}{l}
\eto{\wx_k\nm x_k}-1=
-hD_k^{-1}\Big(1+g^2\eto{x_k\nm\wx_{k-1}}\Big)\\ \\
\eto{\wip_k\nm p_k} = 
\displaystyle\frac{\bigg(1-hD_k^{-1}g^2\eto{x_k\nm\wx_{k-1}}\bigg)}
{\bigg(1-hD_{k+1}^{-1}g^2\eto{x_{k+1}\nm\wx_k}\bigg)}
\end{array}\right.
\end{equation}
where
\begin{equation}
D_k=\eto{p_k}+\delta\Big(1+g^2\eto{x_k\nm\wx_{k-1}}\Big)=\eto{p_k}+O(\delta)
\end{equation}
Lagrangian form of equations of motion:
\begin{equation}
\left\{\begin{array}{l}
h\eto{p_k} = 
-\,\displaystyle\frac{\bigg(\delta\Big(\eto{\wx_k\nm x_k}-1\Big)+h\bigg)}
{\bigg(\eto{\wx_k\nm x_k}-1\bigg)}\;\bigg(1+g^2\eto{x_k\nm\wx_{k-1}}\bigg)\\ \\
h\eto{\wip_k} =
-\,\displaystyle\frac{\bigg(\delta\Big(\eto{\wx_k\nm x_k}-1\Big)+h\bigg)}
{\bigg(\eto{\wx_k\nm x_k}-1\bigg)}\;\bigg(1+g^2\eto{x_{k+1}\nm\wx_k}\bigg)\;
\displaystyle\frac{\bigg(1+g^2\eto{\wx_k\nm\wx_{k-1}}\bigg)}
{\bigg(1+g^2\eto{\wx_{k+1}\nm\wx_k}\bigg)}
\end{array}\right.
\end{equation}
Newtonian form of equations of motion:
\begin{equation}\label{dRTL- gq New}
\displaystyle\frac
{\bigg(\eto{\wx_k\nm x_k}-1\bigg)\,
\bigg(\delta\Big(e\ueto{x_k\nm\undertilde{x_k}}-1\Big)+h\bigg)}
{\bigg(e\ueto{x_k\nm\undertilde{x_k}}-1\bigg)\,
\bigg(\delta\Big(\eto{\wx_k\nm x_k}-1\Big)+h\bigg)}
= \displaystyle\frac
{\bigg(1+g^2\eto{x_{k+1}\nm x_k}\bigg)\,
\bigg(1+g^2\eto{x_k\nm\wx_{k-1}}\bigg)}
{\bigg(1+g^2e\ueto{\,\undertilde{x_{k+1}}\nm x_k}\bigg)\,
\bigg(1+g^2\eto{x_k\nm x_{k-1}}\bigg)}
\end{equation}

\paragraph*{Special case $\delta=h$.} If the parameter $\delta$ is equal
to the (small) stepsize $h$, the map dRTL$-$ simplifies, providing
discretization of the relativistic Toda lattice (\ref{RTL New}) different
from (\ref{dRTL New}). In this case the ''Hamiltonian'' equations of motion
read:
\begin{equation}
\left\{\begin{array}{l}
\eto{\wx_k\nm x_k}=\displaystyle\frac{1}
{1+h\eto{-p_k}\bigg(1+g^2\eto{x_k-\wx_{k-1}}\bigg)}\\ \\
\eto{\wip_k\nm p_k} = 
\displaystyle\frac
{\bigg(1+hg^2\Big(\eto{p_{k+1}}+h\Big)^{-1}\eto{x_{k+1}\nm\wx_k}\bigg)}
{\bigg(1+hg^2\Big(\eto{p_k}+h\Big)^{-1}\eto{x_k\nm\wx_{k-1}}\bigg)}
\end{array}\right.
\end{equation}
The Lagrangian equations of motion read:
\begin{equation}
\left\{\begin{array}{l}
\eto{p_k}=-\, \displaystyle\frac{h\bigg(1+g^2\eto{x_k\nm\wx_{k-1}}\bigg)}
{\bigg(1-\eto{-\wx_k\np x_k}\bigg)}\\ \\
\eto{\wip_k} =
-\,\displaystyle\frac{h\bigg(1+g^2\eto{\wx_k\nm\wx_{k-1}}\bigg)}
{\bigg(1-\eto{-\wx_k\np x_k}\bigg)}\:
\displaystyle\frac{\bigg(1+g^2\eto{x_{k+1}\nm\wx_k}\bigg)}
{\bigg(1+g^2\eto{\wx_{k+1}\nm\wx_k}\bigg)}
\end{array}\right.
\end{equation}
The Newtonian equations of motion:
\begin{equation}\label{dRTL- gq New spec}
\displaystyle\frac{\bigg(1-e\ueto{-\wx_k\np x_k}\bigg)}
{\bigg(1-e\ueto{-x_k\np\undertilde{x_k}}\bigg)}=
\displaystyle\frac{\bigg(1+g^2\eto{x_{k+1}\nm x_k}\bigg)\,
\bigg(1+g^2\eto{x_k\nm\wx_{k-1}}\bigg)}
{\bigg(1+g^2e\ueto{\,\undertilde{x_{k+1}}\nm x_k}\bigg)\,
\bigg(1+g^2\eto{x_k\nm x_{k-1}}\bigg)}
\end{equation}
The system resembles very much the previous discretization of the 
relativistic Toda lattice (\ref{dRTL New}), however the relation between 
the two is far from trivial.

\subsection{Lax representations and $r$--matrix structure}
The Lax representations for the four systems considered in this Section 
may be given in terms of the following Lax matrix:
\begin{equation}\label{RTL gq L}
L_k=\left(\begin{array}{cc}
\lambda\eto{p_k}-\lambda^{-1} & \eto{x_k}\\ \\
-g^2\Big(\eto{p_k}+\delta\Big)\eto{-x_k} & \lambda\delta g^2\end{array}\right)
\end{equation}
which satisfies the $r$--matriz ansatz (\ref{rmat 2 ans}) with the $r$--matrix
(\ref{r}).

The Lax representation for the flow RTL+: (\ref{Lax}) with
\begin{equation}
M_k=\left(\begin{array}{cc}
\lambda^{-2}+g^2\Big(\eto{p_{k-1}}+\delta\Big)\eto{x_k\nm x_{k-1}} & 
-\lambda^{-1}\eto{x_k}\\ \\ 
\lambda^{-1}g^2\Big(\eto{p_{k-1}}+\delta\Big)\eto{-x_{k-1}} & 0
\end{array}\right)
\end{equation}
Lax representation for the map dRTL+: (\ref{dLax}) with 
\begin{equation}
V_k=\left(\begin{array}{cc}
1+h\lambda^{-2}+hg^2\Big(\eto{\wip_{k-1}}+\delta\Big)\eto{x_k\nm\wx_{k-1}} &
 -h\lambda^{-1}\eto{x_k}\\ \\ 
h\lambda^{-1}g^2\Big(\eto{\wip_{k-1}}+\delta\Big)\eto{-\wx_{k-1}} & 1
\end{array}\right)
\end{equation}
Lax representaion for the flow RTL$-$: (\ref{Lax}) with 
\begin{equation}
M_k=\displaystyle\frac{d_k^{-1}}{1+\delta\lambda^2}
\left(\begin{array}{cc}
-\lambda^2\Big(\eto{p_k}+\delta\Big) & -\lambda\eto{x_k}\\ \\
\lambda g^2\Big(\eto{p_k}+\delta\Big)\eto{-x_{k-1}} & 
g^2\eto{x_k\nm x_{k-1}}\end{array}\right)
\end{equation}
Lax representation for the map dRTL$-$: (\ref{dLax inv}) with
\begin{equation}
W_k=I+\displaystyle\frac{hD_k^{-1}}{1+(\delta-h)\lambda^2}
\left(\begin{array}{cc}
\lambda^2\Big(\eto{p_k}+\delta\Big) & \lambda\eto{x_k}\\ \\
-\lambda g^2\Big(\eto{p_k}+\delta\Big)\eto{-\wx_{k-1}} & 
-g^2\eto{x_k\nm \wx_{k-1}}\end{array}\right)
\end{equation}

\paragraph*{Special case $\delta=h$.} In this case, for the map dRTL$-$, 
the dependence of the matrix $W_k$ on $\lambda$ simplifies, because the 
denominator $1+(\delta-h)\lambda^2$ becomes equal to 1.

\setcounter{equation}{0}
\section{Systems RTL$\pm$ and dRTL$\pm$:\newline
The second mixed parametrization}

We consider the following canonical parametrization of the variables $(c,d)$:
\begin{equation}\label{RTL gl par}
d_k=\displaystyle\frac{\eto{\epsilon p_k}-1}{\epsilon}-\eto{x_k\nm x_{k-1}},
\qquad c_k=\eto{x_{k+1}\nm x_k\np \epsilon p_k}
\end{equation}
By small $\epsilon$ it serves as a deformation of (\ref{RTL l par}). The 
corresponding Poisson bracket is $\{\cdot,\cdot\}_1+\epsilon\{\cdot,\cdot\}_2$.
the results of this Section allow to recover that of the section 9 
in the limit $\epsilon\to 0$ (in order to reproduce the Lax representations 
of the Sect.9, the spectral parameter $\lambda$ of this Section
has to be replaced by $1+\epsilon\lambda/2+O(\epsilon^2)$ before perforimg the 
limit $\epsilon\to 0$).

\subsection{The flow RTL+}
Hamilton function:
\begin{equation}
H(x,p)=\epsilon^{-1}H_2^{(+)}(c,d)-\epsilon^{-1}\sum_kp_k
=\sum_k\displaystyle\frac{\eto{\epsilon p_k}-1-\epsilon p_k}{\epsilon^2}+
\sum_k\displaystyle\frac{\eto{\epsilon p_k}-1}{\epsilon}\,\eto{x_{k+1}\nm x_k}
\end{equation}
Hamiltonian equations of motion:
\begin{equation}\label{RTL+ gl Ham}
\left\{\begin{array}{l}
\dot{x}_k=\displaystyle\frac{\Big(\eto{\epsilon p_k}-1\Big)}{\epsilon}+
\eto{x_{k+1}\nm x_k\np\epsilon p_k}\\ \\
\dot{p}_k=
\displaystyle\frac{\Big(\eto{\epsilon p_k}-1\Big)}{\epsilon}\,
\eto{x_{k+1}\nm x_k}
-\displaystyle\frac{\Big(\eto{\epsilon p_{k-1}}-1\Big)}{\epsilon}\,
\eto{x_k\nm x_{k-1}}
\end{array}\right.
\end{equation}
Lagrangian equations of motion:
\begin{equation}\label{RTL+ gl Lagr}
\left\{\begin{array}{l}
\eto{\epsilon p_k}=\displaystyle\frac{(1+\epsilon\dot{x}_k)}
{\Big(1+\epsilon\eto{x_{k+1}\nm x_k}\Big)}\\ \\
\dot{p}_k=\displaystyle\frac{\Big(\dot{x}_k-\eto{x_{k+1}\nm x_k}\Big)}
{\Big(1+\epsilon\eto{x_{k+1}\nm x_k}\Big)}\,\eto{x_{k+1}\nm x_k}-
\displaystyle\frac{\Big(\dot{x}_{k-1}-\eto{x_k\nm x_{k-1}}\Big)}
{\Big(1+\epsilon\eto{x_k\nm x_{k-1}}\Big)}\,\eto{x_k\nm x_{k-1}\Big)}
\end{array}\right.
\end{equation}
The corresponding Newtonian equations of motion read:
\begin{equation}\label{RTL+ gl New}
\ddot{x}_k= 
(1+\epsilon\dot{x}_k)\left(
\displaystyle\frac{\Big(\dot{x}_{k+1}-\eto{x_{k+1}\nm x_k}\Big)}
{\Big(1+\epsilon\eto{x_{k+1}\nm x_k}\Big)}\,\eto{x_{k+1}\nm x_k}-
\displaystyle\frac{\Big(\dot{x}_{k-1}-\eto{x_k\nm x_{k-1}}\Big)}
{\Big(1+\epsilon\eto{x_k\nm x_{k-1}}\Big)}\,\eto{x_k\nm x_{k-1}}\right)
\end{equation}

\subsection{The map dRTL+}

''Hamiltonian'' equations of motion:
\begin{equation}
\left\{\begin{array}{l}
\eto{\wx_k\nm x_k}-1=h\displaystyle\frac{\Big(\eto{\epsilon\wip_k}-1\Big)}
{\epsilon}+
\displaystyle\frac{h\eto{x_{k+1}\nm\wx_k\np\epsilon\wip_k}}
{\bigg(1-h\eto{x_{k+1}\nm\wx_k}\bigg)}\\ \\
\eto{\epsilon\wip_k\nm\epsilon p_k}=
\displaystyle\frac
{\bigg(1+h\Big(\eto{\epsilon \wip_k}-1\Big)\eto{x_{k+1}-\wx_k}\bigg)}
{\bigg(1+h\Big(\eto{\epsilon \wip_{k-1}}-1\Big)\eto{x_k-\wx_{k-1}}\bigg)}
\end{array}\right.
\end{equation}
Lagrangian equations of motion:
\begin{equation}
\left\{\begin{array}{l}
h\eto{\epsilon p_k} = \bigg(\epsilon\Big(\eto{\wx_k\nm x_k}-1\Big)+h\bigg)\;
\displaystyle\frac{\bigg(1-h\eto{x_k\nm \wx_{k-1}}\bigg)}
{\bigg(1+(\epsilon-h)\eto{x_k\nm\wx_{k-1}}\bigg)}\:
\displaystyle\frac{\bigg(1+\epsilon\eto{x_k\nm x_{k-1}}\bigg)}
{\bigg(1+\epsilon\eto{x_{k+1}\nm x_k}\bigg)}\\ \\
h\eto{\epsilon\wip_k} =\bigg(\epsilon\Big(\eto{\wx_k\nm x_k}-1\Big)+h\bigg)
\displaystyle\frac{\bigg(1-h\eto{x_{k+1}\nm\wx_{k}}\bigg)}
{\bigg(1+(\epsilon-h)\eto{x_{k+1}\nm\wx_k}\bigg)}
\end{array}\right.
\end{equation}
Newtonian equations of motion:
\begin{equation}\label{dRTL+ gl New}
\displaystyle\frac{\epsilon\bigg(e\ueto{\wx_k\nm x_k}-1\bigg)+h}
{\epsilon\bigg(e\ueto{x_k\nm\undertilde{x_k}}-1\bigg)+h} 
= \displaystyle\frac{\bigg(1+\epsilon \eto{x_{k+1}\nm x_k}\bigg)\,
\bigg(1-he\ueto{\,\undertilde{x_{k+1}}\nm x_k}\bigg)\,
\bigg(1+(\epsilon-h)\eto{x_k\nm\wx_{k-1}}\bigg)}
{\bigg(1+\epsilon \eto{x_k\nm x_{k-1}}\bigg)\,
\bigg(1-h\eto{x_{k}\nm\wx_{k-1}}\bigg)\,
\bigg(1+(\epsilon-h)e\ueto{\,\undertilde{x_{k+1}}\nm x_k}\bigg)}
\end{equation}

\paragraph*{Special case $\epsilon=h$.} If the parameter $\epsilon$ is
equal to the (small) stepsize $h$, then the equations above greatly
simplify, delivering another discretization of the system (\ref{RTL+ l New}).
In this case we obtain the following ''Hamiltonian'' equations of motion:
\begin{equation}
\left\{\begin{array}{l}
\eto{\wx_k\nm x_k}=\displaystyle\frac{\eto{h\wip_k}}
{\bigg(1-h\eto{x_{k+1}\nm\wx_k}\bigg)}\\ \\
\eto{h\wip_k\nm hp_k}=
\displaystyle\frac{\bigg(1+h\Big(\eto{h\wip_k}-1\Big)\eto{x_{k+1}-\wx_k}\bigg)}
{\bigg(1+h\Big(\eto{h\wip_{k-1}}-1\Big)\eto{x_k-\wx_{k-1}}\bigg)}
\end{array}\right.
\end{equation}
Lagrangian equations of motion:
\begin{equation}
\left\{\begin{array}{l}
\eto{hp_k} = \eto{\wx_k\nm x_k}\;\bigg(1-h\eto{x_k\nm \wx_{k-1}}\bigg)
\displaystyle\frac{\bigg(1+h\eto{x_k\nm x_{k-1}}\bigg)}
{\bigg(1+h\eto{x_{k+1}\nm x_k}\bigg)}\\ \\
\eto{h\wip_k} =\eto{\wx_k\nm x_k}\;
\bigg(1-h\eto{x_{k+1}\nm\wx_{k}}\bigg)
\end{array}\right.
\end{equation}
Newtonian equations of motion:
\begin{equation}\label{dRTL+ gl New spec}
e\ueto{\wx_k\nm 2x_k\np\undertilde{x_k}}
= \displaystyle\frac{\bigg(1+h\eto{x_{k+1}\nm x_k}\bigg)\,
\bigg(1-he\ueto{\,\undertilde{x_{k+1}}\nm x_k}\bigg)}
{\bigg(1+h\eto{x_k\nm x_{k-1}}\bigg)\,
\bigg(1-h\eto{x_{k}\nm\wx_{k-1}}\bigg)}
\end{equation}

\subsection{The flow RTL$-$}

Hamilton function:
\begin{equation}
H(x,p)=H_1^{(-)}(c,d)+\epsilon\sum_kp_k=-\sum_k\log(d_k)+\epsilon\sum_kp_k
\end{equation}
where, recall,
\begin{equation}
d_k=\displaystyle\frac{\eto{\epsilon p_k}-1}{\epsilon}-\eto{x_k\nm x_{k-1}}=
p_k-\eto{x_k\nm x_{k-1}}+O(\epsilon)
\end{equation}
Hamiltonian equations of motion:
\begin{equation}\label{RTL- gl Ham}
\left\{\begin{array}{l}
\dot{x}_k=-d_k^{-1}\;\Big(1+\epsilon\eto{x_k\nm x_{k-1}}\Big)\\ \\
\dot{p}_k=d_{k+1}^{-1}\;\eto{x_{k+1}\nm x_{k}}-
d_k^{-1}\;\eto{x_k\nm x_{k-1}}
\end{array}\right.
\end{equation}
Lagrangian equations of motion:
\begin{equation}\label{RTL- gl Lagr}
\left\{\begin{array}{l}
\eto{\epsilon p_k}=\Big(1+\epsilon\eto{x_k\nm x_{k-1}}\Big)\;
\displaystyle\frac{(\dot{x}_k-\epsilon)}{\dot{x}_k}\\ \\
\dot{p}_k=-\dot{x}_{k+1}\,\displaystyle\frac{\eto{x_{k+1}\nm x_k}}
{\Big(1+\epsilon\eto{x_{k+1}\nm x_k}\Big)}+\dot{x}_k\,\displaystyle\frac
{\eto{x_k\nm x_{k-1}}}{\Big(1+\epsilon\eto{x_k\nm x_{k-1}}\Big)}
\end{array}\right.
\end{equation}
Newtonian equations of motion:
\begin{equation}\label{RTL- gl New}
\ddot{x}_k =
-\dot{x}_k(\dot{x}_k-\epsilon)\left(
\dot{x}_{k+1}\,\displaystyle\frac{\eto{x_{k+1}\nm x_k}}
{\Big(1+\epsilon\eto{x_{k+1}\nm x_k}\Big)}-
\dot{x}_{k-1}\,\displaystyle\frac{\eto{x_k\nm x_{k-1}}}
{\Big(1+\epsilon\eto{x_k\nm x_{k-1}}\Big)}\right)
\end{equation}

\subsection{The map dRTL$-$}

''Hamiltonian'' equations of motion:
\begin{equation}
\left\{\begin{array}{l}
\eto{\wx_k\nm x_k}-1=
-hD_k^{-1}\Big(1+\epsilon\eto{x_k\nm\wx_{k-1}}\Big)\\ \\
\eto{\epsilon\wip_k\nm\epsilon p_k} = 
\displaystyle\frac{\bigg(1-h\epsilon D_k^{-1}\eto{x_k\nm\wx_{k-1}}\bigg)}
{\bigg(1-h\epsilon D_{k+1}^{-1}\eto{x_{k+1}\nm\wx_k}\bigg)}
\end{array}\right.
\end{equation}
where
\begin{equation}
D_k=\displaystyle\frac{\eto{\epsilon p_k}-1}{\epsilon}-\eto{x_k\nm\wx_{k-1}}
=p_k-\eto{x_k\nm\wx_{k-1}}+O(\epsilon)
\end{equation}
Lagrangian form of equations of motion:
\begin{equation}
\left\{\begin{array}{l}
\eto{\epsilon p_k} = 
\displaystyle\frac{\bigg(\eto{\wx_k\nm x_k}-1-h\epsilon\bigg)}
{\bigg(\eto{\wx_k\nm x_k}-1\bigg)}\;
\bigg(1+\epsilon\eto{x_k\nm\wx_{k-1}}\bigg)\\ \\
\eto{\epsilon\wip_k} =
\displaystyle\frac{\bigg(\eto{\wx_k\nm x_k}-1-h\epsilon\bigg)}
{\bigg(\eto{\wx_k\nm x_k}-1\bigg)}\;\bigg(1+\epsilon\eto{x_{k+1}\nm\wx_k}\bigg)\;
\displaystyle\frac{\bigg(1+\epsilon\eto{\wx_k\nm\wx_{k-1}}\bigg)}
{\bigg(1+\epsilon\eto{\wx_{k+1}\nm\wx_k}\bigg)}
\end{array}\right.
\end{equation}
Newtonian form of equations of motion:
\begin{equation}\label{dRTL- gl New}
\displaystyle\frac
{\bigg(\eto{\wx_k\nm x_k}-1\bigg)\,
\bigg(e\ueto{x_k\nm\undertilde{x_k}}-1-h\epsilon\bigg)}
{\bigg(e\ueto{x_k\nm\undertilde{x_k}}-1\bigg)\,
\bigg(\eto{\wx_k\nm x_k}-1-h\epsilon\bigg)}
= \displaystyle\frac
{\bigg(1+\epsilon \eto{x_{k+1}\nm x_k}\bigg)\,
\bigg(1+\epsilon \eto{x_k\nm\wx_{k-1}}\bigg)}
{\bigg(1+\epsilon e\ueto{\,\undertilde{x_{k+1}}\nm x_k}\bigg)\,
\bigg(1+\epsilon \eto{x_k\nm x_{k-1}}\bigg)}
\end{equation}

\subsection{Lax representations and $r$--matrix structure}
The Lax representations for the four systems considered in this Section 
may be given in terms of the following Lax matrix:
\begin{equation}\label{RTL gl L}
L_k=\left(\begin{array}{cc}
\lambda\eto{\epsilon p_k}-\lambda^{-1} & \epsilon\eto{x_k}\\ \\
-\Big(\eto{\epsilon p_k}-1\Big)\eto{-x_k} & -\epsilon\lambda
\end{array}\right)
\end{equation}
which satisfies the $r$--matriz ansatz (\ref{rmat eps ans}) with the $r$--matrix
(\ref{r}).

The Lax representation for the flow RTL+: (\ref{Lax}) with
\begin{equation}
M_k=\left(\begin{array}{cc}
\displaystyle\frac{\lambda^{-2}-1}{\epsilon}+
\Big(\eto{\epsilon p_{k-1}}-1\Big)\eto{x_k\nm x_{k-1}} & 
-\lambda^{-1}\eto{x_k}\\ \\ 
\lambda^{-1}\,\displaystyle\frac{\Big(\eto{\epsilon p_{k-1}}-1\Big)}{\epsilon}
\,\eto{-x_{k-1}}
 & 0\end{array}\right)
\end{equation}
Lax representation for the map dRTL+: (\ref{dLax}) with 
\begin{equation}
V_k=\left(\begin{array}{cc}
1-\displaystyle\frac{h}{\epsilon}+\displaystyle\frac{h}{\epsilon}\lambda^{-2}
+\displaystyle\frac{h(\epsilon-h)}{\epsilon}
\Big(\eto{\epsilon\wip_{k-1}}-1\Big)\eto{x_k\nm\wx_{k-1}} &
 -h\lambda^{-1}\eto{x_k}\\ \\ 
\displaystyle\frac{h}{\epsilon}\lambda^{-1}
\Big(\eto{\epsilon\wip_{k-1}}-1\Big)\eto{-\wx_{k-1}} & 1
\end{array}\right)
\end{equation}
Lax representaion for the flow RTL$-$: (\ref{Lax}) with 
\begin{equation}
M_k=\displaystyle\frac{d_k^{-1}}{\Big(\lambda^2-1\Big)}
\left(\begin{array}{cc}
\lambda^2\Big(\eto{\epsilon p_k}-1\Big) & \lambda\epsilon\eto{x_k}\\ \\
-\lambda\Big(\eto{\epsilon p_k}-1\Big)\eto{-x_{k-1}} & 
-\epsilon\eto{x_k\nm x_{k-1}}\end{array}\right)
\end{equation}
Lax representation for the map dRTL$-$: (\ref{dLax inv}) with
\begin{equation}
W_k=I+\displaystyle\frac{hD_k^{-1}}{\Big(1-(1+h\epsilon)\lambda^2\Big)}
\left(\begin{array}{cc}
\lambda^2\Big(\eto{\epsilon p_k}-1\Big) & \lambda\epsilon\eto{x_k}\\ \\
-\lambda \Big(\eto{\epsilon p_k}-1\Big)\eto{-\wx_{k-1}} & 
-\epsilon\eto{x_k\nm \wx_{k-1}}\end{array}\right)
\end{equation}

\paragraph*{Special case $\epsilon=h$.} In this case the Lax representation
for the map dRTL+ simplifies significantly: it reads (\ref{dLax}) with
\begin{equation}\label{RTL gl L spec}
L_k=\left(\begin{array}{cc}
\lambda\eto{hp_k}-\lambda^{-1} & h\eto{x_k}\\ \\
-\Big(\eto{hp_k}-1\Big)\eto{-x_k} & -h\lambda
\end{array}\right)
\end{equation}
\begin{equation}
V_k=\left(\begin{array}{cc}
\lambda^{-2} &
 -h\lambda^{-1}\eto{x_k}\\ \\ 
\lambda^{-1}\Big(\eto{h\wip_{k-1}}-1\Big)\eto{-\wx_{k-1}} & 1
\end{array}\right)
\end{equation}

\section{Bibliographical remarks}
The Toda lattice (\ref{Toda New}) was discovered by Toda, see:
\begin{itemize}
\item M.Toda. Theory of nonlinear lattices. Springer, 1981.
\end{itemize}
The relativistic Toda lattice (\ref{RTL New}) was invented by Ruijsenaars:
\begin{itemize}
\item S.N.M.Ruijsenaars. Relativistic Toda systems. {\it Commun. Math. Phys.}, 
{\bf 133} (1990) 217--247.
\end{itemize}
The modified Toda lattice (\ref{mToda New}) seems to appear for the first time 
in the work by Yamilov:
\begin{itemize}
\item R.I.Yamilov. Classification of Toda--type scalar lattices. In: {\it
Nonlinear Evolution Equations and Dynamical Systems, Proc. of the 8th 
International Workshop}, World Scientific Publishing, 1993, pp. 423--431
(translation of the preprint of 1989 in Russian).
\end{itemize}
The Newtonian continuous time systems (\ref{RTL+ l New}), (\ref{RTL- l New})
were introduced in:
\begin{itemize}
\item Yu.B.Suris. New integrable systems related to the relativistic Toda 
lattice. {\it J. Phys. A: Math. and Gen.} {\bf 30} (1997) 1745--1761.
\end{itemize}
and their generalizations (\ref{RTL+ gq New}), (\ref{RTL- gq New}),
(\ref{RTL+ gl New}), (\ref{RTL- gl New}) seem to be introduced in the present
paper for the first time.

The $2\times 2$ Lax representation for the Toda lattice with the matrices
(\ref{Toda L}), (\ref{Toda M}) belongs to Faddeev with co-authors, see:
\begin{itemize}
\item L.D.Faddeev, L.A.Takhtajan. Hamiltonian methods in the theory of solitons.
\newline Springer, 1987.
\end{itemize}
Analogous representations for the two flows of the relativistic Toda lattice
with the matrices (\ref{rToda L}), (\ref{rToda+ M}), (\ref{rToda- M})
 appeared for the first time in:
\begin{itemize}
\item Yu.B.Suris. Discrete time generalized Toda lattices: complete 
integrability and relation with relativistic Toda lattices. {\it Phys. Lett. A}
{\bf 145} (1990) 113--119.
\end{itemize}
For the modified Toda lattice (\ref{mToda New}) and for the flow 
(\ref{RTL+ l New}) the $2\times 2$ Lax representations were derived by 
Deconinck using an analog of the Walquist--Estabrook approach (they are 
equivalent to (\ref{mToda L}), (\ref{mToda M}) and to (\ref{RTL l L}),
(\ref{RTL+ l M}), respectively):
\begin{itemize}
\item B.Deconinck. A constructive test for integrability of semi--discrete
systems. {\it Phys. Lett. A} {\bf 223} (1996) 45--54.
\end{itemize}
For the flow (\ref{RTL- l New}), as well as for the flows from the 
Sect. 10, 11, such Lax representations are introduced here for the first time.

Turning to the discrete time systems. The Hirota's discretization of the Toda
lattice (Sect. 7.1) was found in:
\begin{itemize}
\item R.Hirota. Nonlinear partial difference equations. II. Discrete time
Toda equation. {\it J. Phys. Soc. Japan} {\bf 43} (1977) 2074--2078; III
B\"acklund transformations for the discrete time Toda equations. {\it J. Phys. 
Soc. Japan} {\bf 45} (1978) 321--332.
\end{itemize}
(However, the Lax representation given in the present paper seems to be new). 
The standard discretization of the Toda lattice (Sect. 7.2), together with
the Lax representation, was found in the cited above paper of 1990 by the 
author, and even slightly earlier in the Russian version of another paper:
\begin{itemize}
\item Yu.B.Suris. Generalized toda chains in discrete time. {\it Leningrad
Math. J.} {\bf 2} (1991) 339--352.
\end{itemize}
The discretizations from the Sect. 4 appeared in:
\begin{itemize}
\item Yu.B.Suris. Bi--Hamiltonian structure of the $qd$ algorithm and new 
discretizations of the Toda lattice. {\it Physics Letters A} {\bf 206}
(1995) 153--161.
\end{itemize}
the discretizations from the Sect. 5, 6, 7.3 -- in:
\begin{itemize}
\item Yu.B.Suris. On some integrable systems related to the Toda lattice. 
{\it J. Phys. A: Math. and Gen.} {\bf 30} (1997)
\end{itemize}
the discretizations from the Sect. 8 -- in:
\begin{itemize}
\item Yu.B.Suris. A discrete--time relativistic Toda lattice. {\it J. Phys. A: 
Math. and Gen.} {\bf 29} (1996) 451--465.
\end{itemize}
the discretizations from the Sect. 9 -- in the already cited paper
\begin{itemize}
\item Yu.B.Suris. New integrable systems related to the relativistic Toda 
lattice. {\it J. Phys. A: Math. and Gen.}, {\bf 30} (1997) 1745--1761.
\end{itemize}
The $2\times 2$ Lax representations for all these discrete time systems seem
to appear in the present paper for the first time, as well as the discrete 
time systems from the Sect. 10, 11.

\end{document}